\documentclass[10pt,a4paper]{article}
\usepackage{jcappub}

\title{Interacting dark energy models in fractal cosmology}

\author{O.\,A.\,Lemets}
\author{D.\,A.\,Yerokhin}
\affiliation{Akhiezer Institute for Theoretical Physics,
National Science Center "Kharkov Institute of Physics and
Technology", Akademicheskaya Str. 1, 61108 Kharkov, Ukraine}
\emailAdd{oleg.lemets@gmail.com}
\emailAdd{denyerokhin@gmail.com}

\abstract{
We investigate interacting dark energy models in the framework of fractal cosmology. We discuss a fractal FRW universe filled with the dark energy and dark matter which interact with each other. We obtain the equation for the relative density of dark matter and dark energy and the deceleration parameter. This model demonstrates new types of evolution, which are not common to cosmological models with this type of interaction.}
\keywords{dark energy; fractal cosmology transient acceleration,scalar field}
\arxivnumber{1202.3457}
\begin{document}
\maketitle
\flushbottom
\toccontinuoustrue
\section{Introduction}

In the present paper, we study interacting dark energy models in the framework of fractal cosmology proposed by Calcagni. In  \cite{Calcagni1,Calcagni2} he proposed a model for a power-counting renormalizable field theory living in a fractal spacetime. The action in this model is Lorentz covariant and equipped with a Stieltjes measure.

For this model we calculate the relative density of dark matter and dark energy, the deceleration parameter and discuss their physical implications via the numerical integration of derived equations of motion.

The motivation of fractal cosmology models is the following\cite{Calcagni3}: first, one can see that most theories of quantum gravity describe the universe as a dimensional flow.Second, one can ask whether and how these properties are connected to the problem of UV divergence. For this intention, geometry and field theory were proposed in \cite{Calcagni4}, in which dimensional flow is an essential property and where UV limitation can be checked explicitly. Thus, these theories are completely independent from different models of quantum gravity.

Fractal features of quantum gravity theories in $D$ dimensions have been investigated for some cases. Renormalizability of perturbative quantum gravity theories at and near two topological dimensions drew mutual concern on $D=2+\epsilon$ models, which are expected to help in better understanding of $D=4$ case \cite{GKT,ChD,KN,Wei79,JJ,KKN,AKNT,KPZ}.

Assuming that matter is minimally coupled with gravity, the total action is \cite{Calcagni1,Calcagni2}
\begin{equation}\label{action}
    S=S_g+S_{\rm m}\,,
\end{equation}
where $S_g$ is
\begin{equation}\label{action_grav}
    S_g=\frac{M_p^2}{2}\int {\rm d}\varrho(x)\,\sqrt{-g}\,\left(R-2\lambda-\omega\partial_\mu v\partial^\mu v\right)\,,
\end{equation}
and
\begin{equation}\label{action_matter}
S_{\rm m}=\int {\rm d}\varrho \sqrt{-g} {\cal L}_{\rm m}
\end{equation}
is the matter action. Here  $g$ is the determinant of the  metric tensor $g_{\mu\nu}$, $M_p^{-2}=8\pi G$ is reduced Planck mass, $R$ is Ricci scalar,  $\lambda$ the bare cosmological constant, and the term proportional to $\omega$ has been added because $v$, similar the other geometric field $g_{\mu\nu}$, is now dynamical. The scaling dimension of $\varrho$ is $[\varrho]=-D\alpha\neq -D$, where $\alpha>0$ is a positive parameter.

 The derivation of the Einstein equations goes almost like in scalar-tensor models. Taking the variation of the action (\ref{action}) with respect to
the Friedmann-Robertson-Walker (FRW) metric $g_{\mu\nu}$, one can obtain the Friedmann equations in a fractal universe as  was found in \cite{Calcagni2}
\begin{equation}\label{fried}
    \left(\frac{D}2-1\right)H^2+H\frac{\dot v}{v}-\frac{1}{2}\frac{\omega}{D-1}\dot v^2=\frac{1}{M_p^2(D-1)}\rho +\frac{\lambda}{D-1}-\frac{k}{a^2}\,,
\end{equation}
\begin{equation}\label{fried2}
    \frac{\Box v}{v}-(D-2)\left(H^2+\dot H-H\frac{\dot v}{v}+\frac{\omega}{D-1}\dot v^2\right)+\frac{2\lambda}{D-1}=\frac{1}{M_p^2(D-1)}\left[(D-3)\rho+(D-1)p\right].
\end{equation}
where $H=\dot{a}/a$ is the Hubble parameter, $\rho$ and $p$ are the total energy density and pressure of the ideal fluid composing the Universe. The parameter $k$ denotes the curvature of the Universe,
where $k=-1, 0 , +1$ for the close, flat and open Universe respectively. As is easy can see when $v=1$, Eqs.(\ref{fried}) and (\ref{fried2}) transform to the standard Friedmann equations in Einstein GR.

If $\rho+p\neq 0$, one can get a purely gravitational constraint (see \cite{Calcagni2} for detail):
\begin{equation}\label{grav_constr}
    \dot{H}+(D-1)H^2+\frac{2k}{a^2}+\frac{\Box v}{v}+H\frac{\dot v}{v}+\omega (v\Box v-\dot v^2)=0\,.
\end{equation}
The continuity equation in fractal cosmology takes the form
\begin{equation}\label{cont_eq}
    \dot\rho+\left[(D-1)H+\frac{\dot v}{v}\right](\rho+p)=0\,,
\end{equation}
When $v=1$ and $D=4$, we recover the standard Friedmann equations in four dimensions, eqs. (\ref{fried}) and (\ref{fried2}) (no gravitational constraint):
\begin{eqnarray}
&&H^2=\frac{1}{3M_p^2}\rho+\frac{\lambda}{3}-\frac{k}{a^2}\,,\label{fr1}\\
&&H^2+\dot{H}=-\frac{1}{6M_p^2}(3p+\rho)+\frac{\lambda}{3}\,.
\end{eqnarray}

On the other hand, for the measure weight
\begin{equation}\label{vt}
v=t^{-\beta},
\end{equation}
where $\beta$ is given by $\beta\equiv D(1-\alpha)$, the gravitational constraint is switched on.  The UV regime, in fact, describes short scales at which inhomogeneities should play some role. If these are small, the modified Friedmann equations define a background for perturbations rather than a self-consistent dynamics.

Recently \cite{Karami}   the holographic, new agegraphic and ghost dark energy models in the framework of fractal cosmology was investigated. In the next section we consider the universe in which dark energy interacting with dark matter.

\section{Dark sector interaction in fractal cosmology}

In this section, we derive the first order differential equations that describe the evolution of interacting dark matter
and dark energy  in the framework of spatially flat $k=0$ fractal cosmology.

\subsection{The general equations}

It is well known that interaction between the components in the Universe must be
introduced in such a way that preserves the covariance of the
energy-momentum tensor $T^{\,\,\mu\nu}_{(tot)\,\,;\nu}=0,$ therefore
$T^{\,\,\mu\nu}_{m\,\,;\nu}=-T^{\,\,\mu\nu}_{x\,\,;\nu}\neq
0,$ where $u_\nu$ is the 4-velocity. The conservation equations in
that case take the form:
\begin{equation}\label{ref1}
 u_\nu T^{\,\,\mu\nu}_{_m\,;\mu}
 =-u_\nu T^{\,\,\mu\nu}_{x\,;\mu}=
 -Q.
\end{equation}
For four-dimensional space with FRW-metric in fractal case and  natural parameterization of the function $v = t^{-\beta}$, equations (\ref{cont_eq}) transform to:
\begin{eqnarray}
  &&\dot{\rho}_m+\left(3H-\beta t^{-1}\right)\rho_m=Q,\label{CeQFrm}\\
 &&\dot{\rho}_{x}+(1+w)\left(3H-{\beta} t^{-1}\right)\rho_{x}=-Q,\label{CeQFrm}
\end{eqnarray}
where $\rho_m$ and $\rho_{x}$ are densities of dark matter and
dark energy respectively, $w$ is the state parameter for dark energy.
It is convenient to use the relative energy densities of dark energy and dark matter  in accordance with standard definitions:
\begin{equation}\label{Rel_den}
\Omega_m = \frac{\rho_m}{3M_p^2H^2},~~ \Omega_x = \frac{\rho_x}{3M_p^2H^2}.
\end{equation}
The above equation can be written in terms of this density parameters as the following:
\begin{equation}\label{OmegaCeQ_sys}
\begin{array}{c}
 \dot{\Omega}_m+\left(3H-\beta t^{-1}\right)\Omega_m+2\Omega_m\frac{\dot{H}}{H}=\frac{Q}{3M_p^2H^2},\\
 \\
\dot{\Omega}_x+(1+w_x)\left(3H-\beta t^{-1}\right)\Omega_x+2\Omega_x\frac{\dot{H}}{H}=-\frac{Q}{3M_p^2H^2},\\
\end{array}
\end{equation}

where the dot denotes  derivative with respect to cosmic time $t.$
The differential equation for the Hubble parameter has the form
\begin{equation}\label{dot_H}
\dot H+H^2-\frac{\beta  H}{2t}+\frac{\beta (\beta +1)}{2t^2}+\frac{\omega  \beta ^2}{3t^{2(\beta +1)}}=-\frac{1}{2}((1+3w)\Omega_x +\Omega_m)H^2.
\end{equation}
In order to obtain the Friedmann equation  in terms of the relative densities it is necessary to enter fictitious density same way as
$\Omega_k=k/(a^2H^2).$  So, we introduce the fractal relative density:
\begin{equation}\label{Rel_denFr}
\Omega_f = \frac{\omega \dot{v}^2}{6H^2}- \frac{\dot{v}}{Hv}.
\end{equation}
Taking into account the ansatz $v=t^{-\beta},$ we obtain the equation of motion for fractal relative density
\begin{equation}\label{Rel_denFr_time}
\Omega_f = \frac{\omega \beta^2}{6H^2t^{2(\beta+1)}}+ \frac{\beta}{Ht},
\end{equation}
Thus, the Friedman equation can be re-written in a very elegant form
\begin{equation}\label{Fr_omega}
\sum_{\alpha = k, f, x, m}\Omega_\alpha  \equiv 1.
\end{equation}
Note that in frames of this definition the values of the relative density  $\Omega_x$ or $\Omega_m$ can exceed 1.

\subsection{Linear interaction of dark matter and dark energy}
In view of the continuity equations, the interaction between dark matter and dark energy must be a function of the energy densities multiplied by a quantity with  dimensions of  inverse time. For the
latter, in the context of cosmology, the obvious choice is the Hubble parameter $H$. So, the
interaction between dark components  could be expressed phenomenologically in such forms as $Q = Q(H (\rho_{x}+\rho_{m}))$, or most generally in the form $Q = Q(H \rho_{x}, H \rho_{m}),$ which
leads to $Q \simeq \delta H\rho_{x} +\gamma H \rho_{m}$ as the first term in its more general expression.
Below we consider the simplest form of interaction -- the linear combination of the decompositions densities of dark matter and dark energy in the flat Friedmann-Robertson-Walker fractal Universe  with:
\begin{equation}\label{Q lin}
Q \equiv H(\delta\rho_{x} + \gamma\rho_{m}).
\end{equation}
In this case, the equations of motion take the form
\begin{eqnarray}
&&\dot{\Omega}_m+\left(3H-\beta t^{-1}\right)\Omega_m+2\Omega_m\frac{\dot{H}}{H} = H(\delta\Omega_{x} + \gamma\Omega_{m}), \nonumber \\
&&\dot{\Omega}_x+(1+w_x)\left(3H-\beta t^{-1}\right)\Omega_x+2\Omega_x\frac{\dot{H}}{H}=-H(\delta\Omega_{x} + \gamma\Omega_{m}), \label{Omega_xsys}\\
&&\dot{\Omega}_f+\left(\frac{\dot{H}}{H}+2(1+\beta) t^{-1}\right)\Omega_f-\frac{(1+2\beta)\beta}{H t} = 0.\nonumber
\end{eqnarray}
Since the equations explicitly depend on time, it is not possible to find their analytical solution.

This is a non-autonomous system of equations that is extremely cumbersome.
  Sometimes it is even more convenient to make it autonomous  introduction of auxiliary variable and thus effective extension of the phase space.

So we introduce the new variables, following to the original work \cite{Copeland_etall}
\begin{equation}\label{xyzuv}
x^2 = \frac{\rho_x}{3M_p^2H^2},~~ y^2 = \frac{\rho_m}{3M_p^2H^2},~~ z^2 = \frac{\omega \beta^2}{6H^2t^{2(\beta+1)}},~~ u^2 = \frac{\beta}{Ht}.
\end{equation}
Note that in this new variables the relative densities of the components take the form
\begin{equation}\label{omega_xyzuv}
\Omega_x = x^2,~~ \Omega_m= y^2,~~ \Omega_f=z^2+ u^2,
\end{equation}
where $\Omega_f$  is fractal relative density as mentioned above. Then Friedman equation becomes
\begin{equation}\label{scalar_frid}
    1=\Omega_x+\Omega_m+\Omega_f=x^2+y^2+z^2+u^2.
\end{equation}
One can see that the variables satisfy the equations
\begin{eqnarray}
 &&x'=-\frac12 x\left[2\frac{H'}{H}+(1+w)(3-u^2)+\delta+\gamma\frac{y^2}{x^2}\right],\nonumber\\
 &&y'=-y\left[2\frac{H'}{H}-u^2+3-\gamma-\delta\frac{x^2}{y^2}\right],\nonumber\\
 &&z'=-z\left[\frac{H'}{H}+z^2\left(1+\frac1\beta\right)\right],\label{xyzuH}\\
 &&u'=-\frac12 u\left[\frac{H'}{H}+\frac{u^2}{\beta}\right],\nonumber\\
 &&\frac{H'}{H}=-1-\frac12 u^2\left(u^2+\frac{1}{\beta}-1\right) -2z^2-\frac{1}{2}\left((1+3w)x^2+y^2\right),\nonumber
\end{eqnarray}
where the prime denotes here $d/dN$ and $N=\ln a$.
The autonomous system of equations have attracted considerable attention since the governing equations can be reduce to a relatively simple finite-dimensional dynamical system, thereby enabling us to study the models standard qualitative techniques.

It is very important to note that when analyzing this system we must take into account the specifics of these \eqref{xyzuv} variables.
It is in the first of all concerns the variables $z$ and $u.$  One can immediately understand how they must behave in various stages of cosmological evolution. In the matter dominated phase the factor behaves as $a(t)\sim t^{-\alpha}$  and the Hubble parameter as $H(t)\sim t^{-1},$ thus $u^2 \simeq const $ and $z^2 \equiv u^2 t^{-2\beta} \simeq const\cdot t^{-2\beta}.$
At the stage of accelerated expansion of the universe, Hubble parameter $H(t)\to const,$  therefore at $t\to \infty$ that is, at the late-time of the universe evolution, we have $\{z, u\} \to 0.$

\subsection{Analyzable case of dark matter and dark energy interaction}

The analytical solution can be found only in the case when the Hubble parameter is inversely proportional to time, which is typical, for example, at the stage  of nonrelativistic matter dominance. Suppose that at this stage the Hubble parameter has the form $H=\sigma t^{-1},$ then the equations \eqref{OmegaCeQ_sys} take the following form

\begin{equation}\label{Omega_m_xsys}
  \begin{array}{c}
   {\Omega}'_m=\theta\Omega_m+\sigma\delta\Omega_{x},\\
    {\Omega}'_x=-\delta\gamma\Omega_{m}+\upsilon\Omega_x,
  \end{array}
\end{equation}
where $\theta=2+\gamma\sigma+\beta-3\sigma, ~\upsilon=2-(1+w)(3\sigma-\beta)-\delta\sigma,$ and    the prime   denotes derivative with respect to logarithm of cosmic time $'\equiv\frac{d}{d\ln t}.$  Note also that the parameter is physically meaningful  under condition $\sigma > 0,$ because we do not consider the collapsing universe.
In this regime of evolution of the Universe the system of equations is autonomous and can  be exactly solved.
Characteristic equation of the system \eqref{Omega_m_xsys} has the form
\begin{equation}\label{Omega_m_xsys_char_eq}
\tau^2-(\theta+\upsilon)\tau+\delta^2\sigma\gamma+\theta\upsilon=0,
\end{equation}
its roots are equal to:
\begin{equation}\label{Omega_m_xsys_roots}
    \tau_{\pm}=\frac{\theta+\upsilon}{2}\left[1\pm\sqrt{1-4\frac{(\delta^2\sigma\gamma+\theta\upsilon)}{(\theta+\upsilon)^2}}\right]
\end{equation}
Let us consider possible types of solutions, and indicate the critical points corresponding to them.
As one can see, this model contains many parameters, making it cumbersome to analyze. Note that due to this feature the system  describes all possible types of critical points typical for coarse equilibrium states.

 Recall that the values of $\beta$ in the IR and UV regimes are $\beta_{\rm
IR}=0$ and $\beta_{\rm UV}=2$, respectively. The UV regime, in fact, describes short scales at which inhomogeneities should play some role. If these are small, the modified Friedmann equations define a background for perturbations rather than a self-consistent dynamics.

\begin{figure}[t]
\centering
\includegraphics[width=0.3\textwidth]{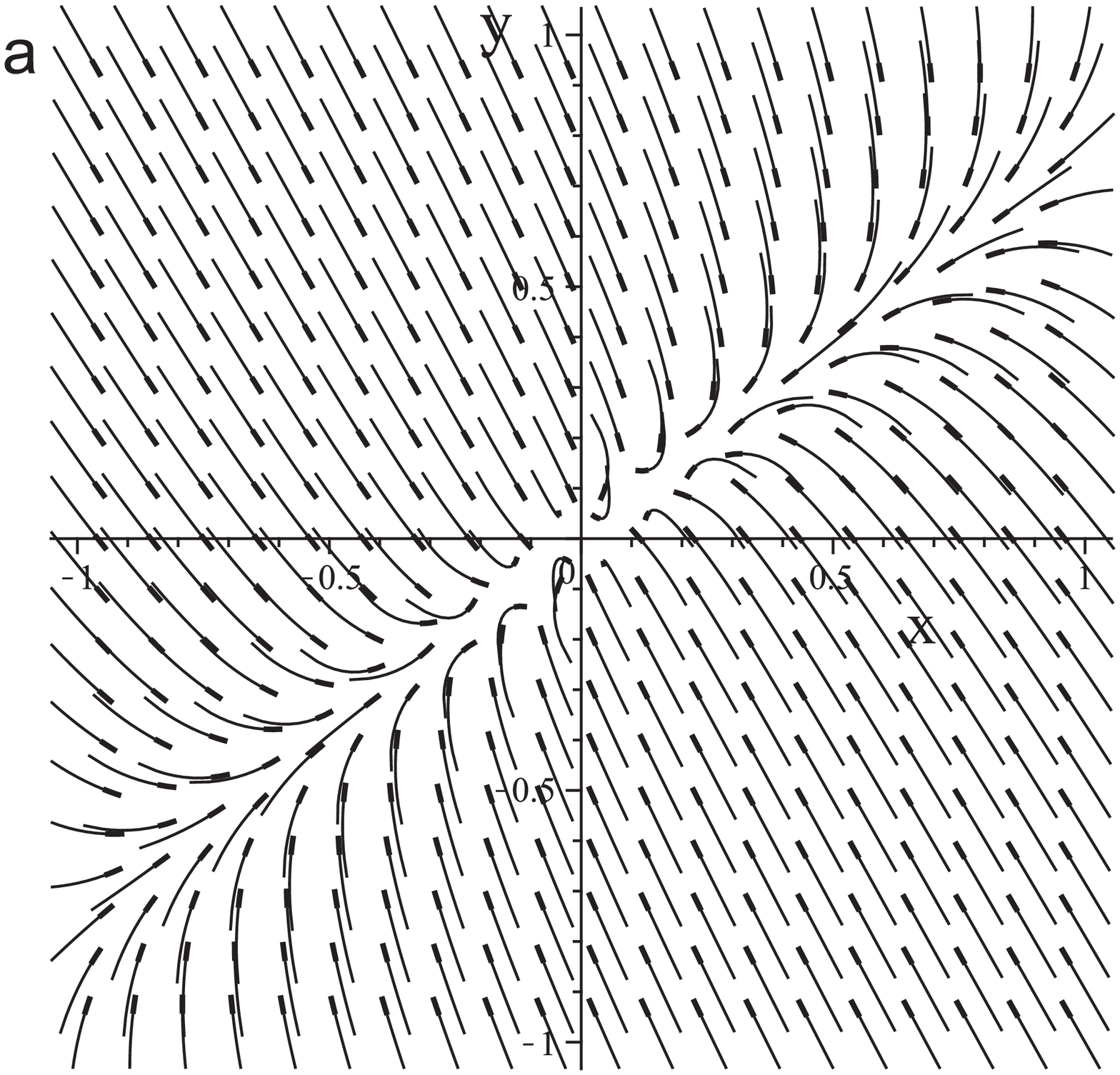}
\includegraphics[width=0.3\textwidth]{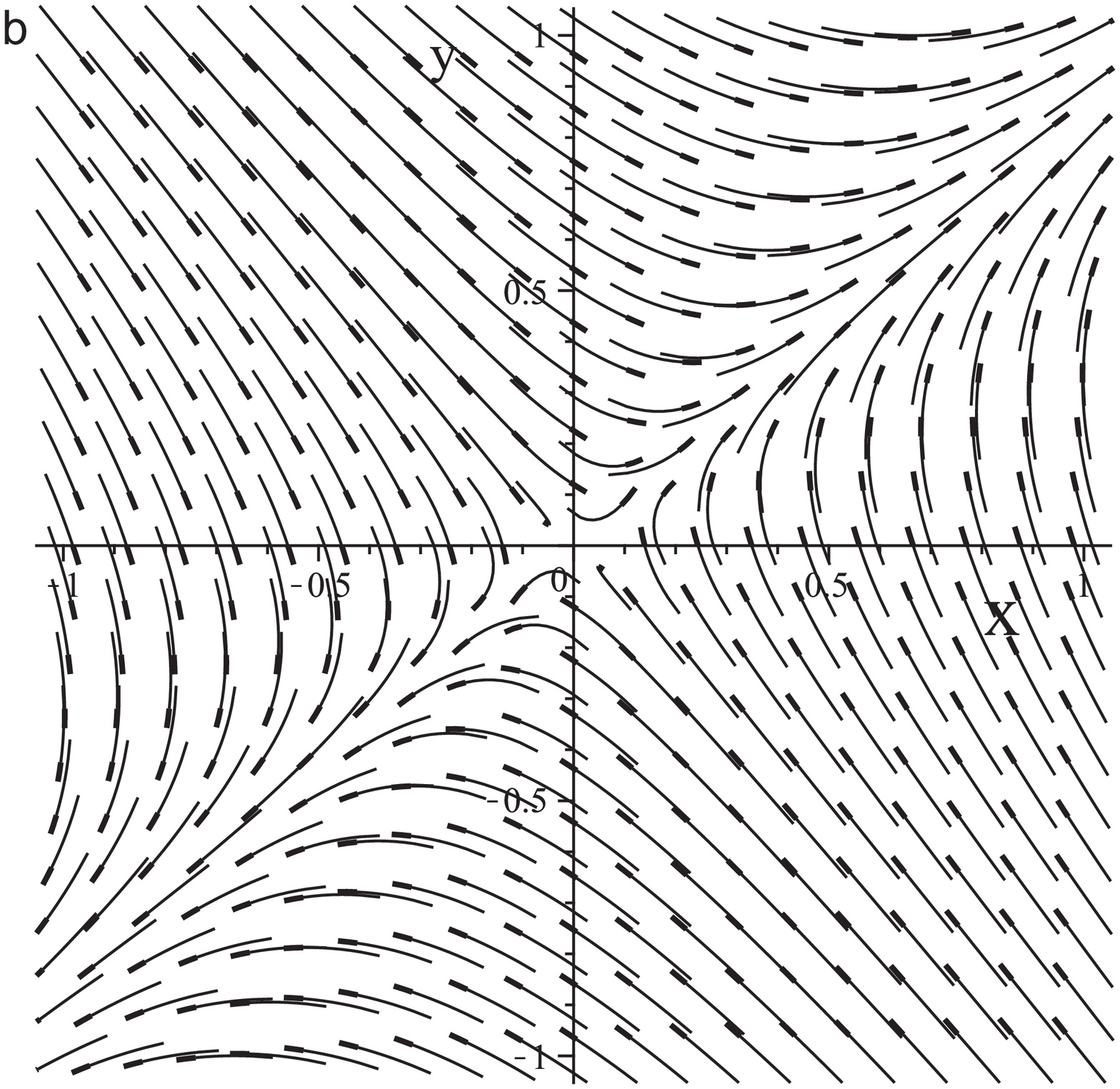}
\includegraphics[width=0.3\textwidth]{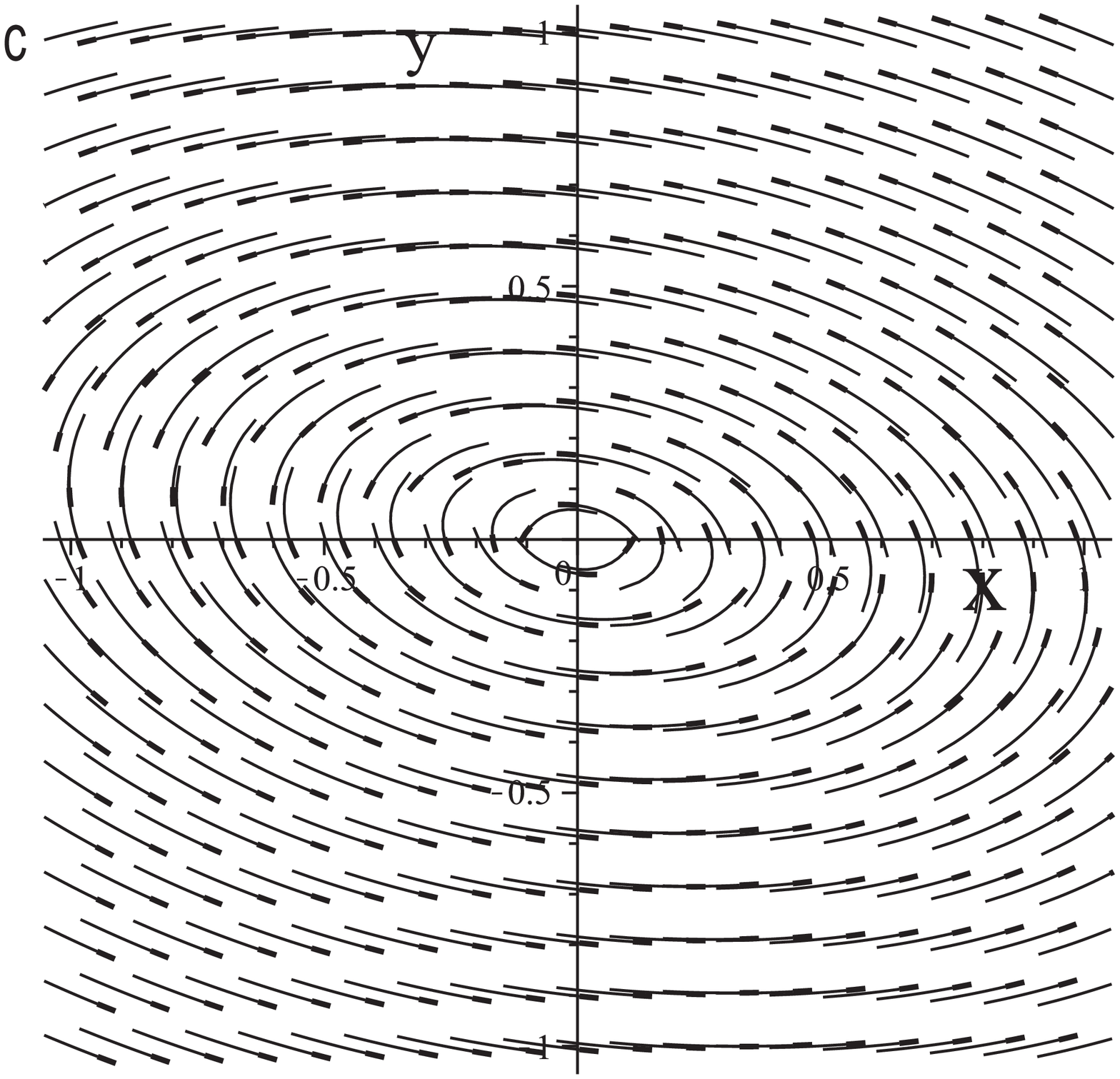}
\includegraphics[width=0.3\textwidth]{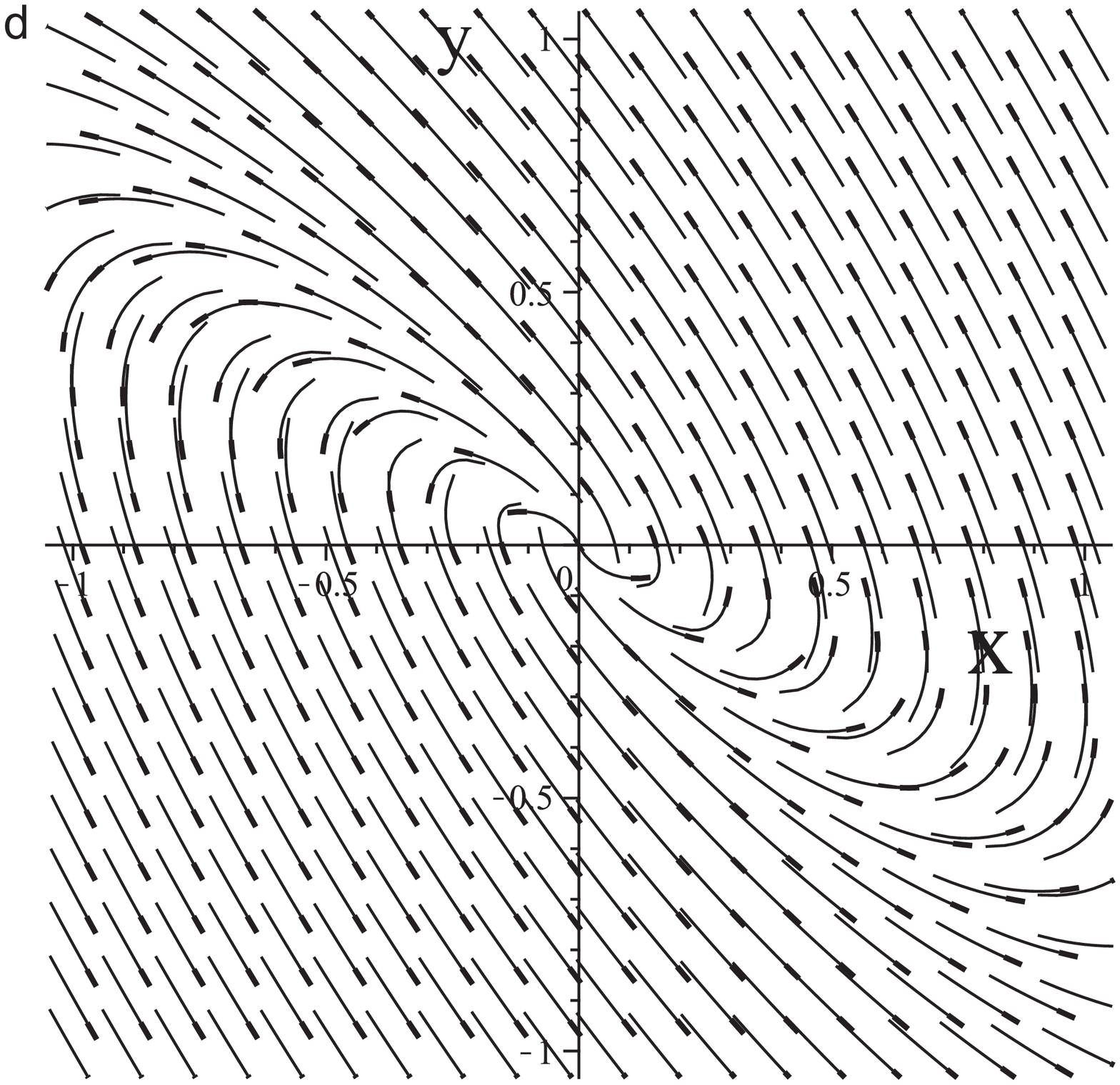}
\includegraphics[width=0.3\textwidth]{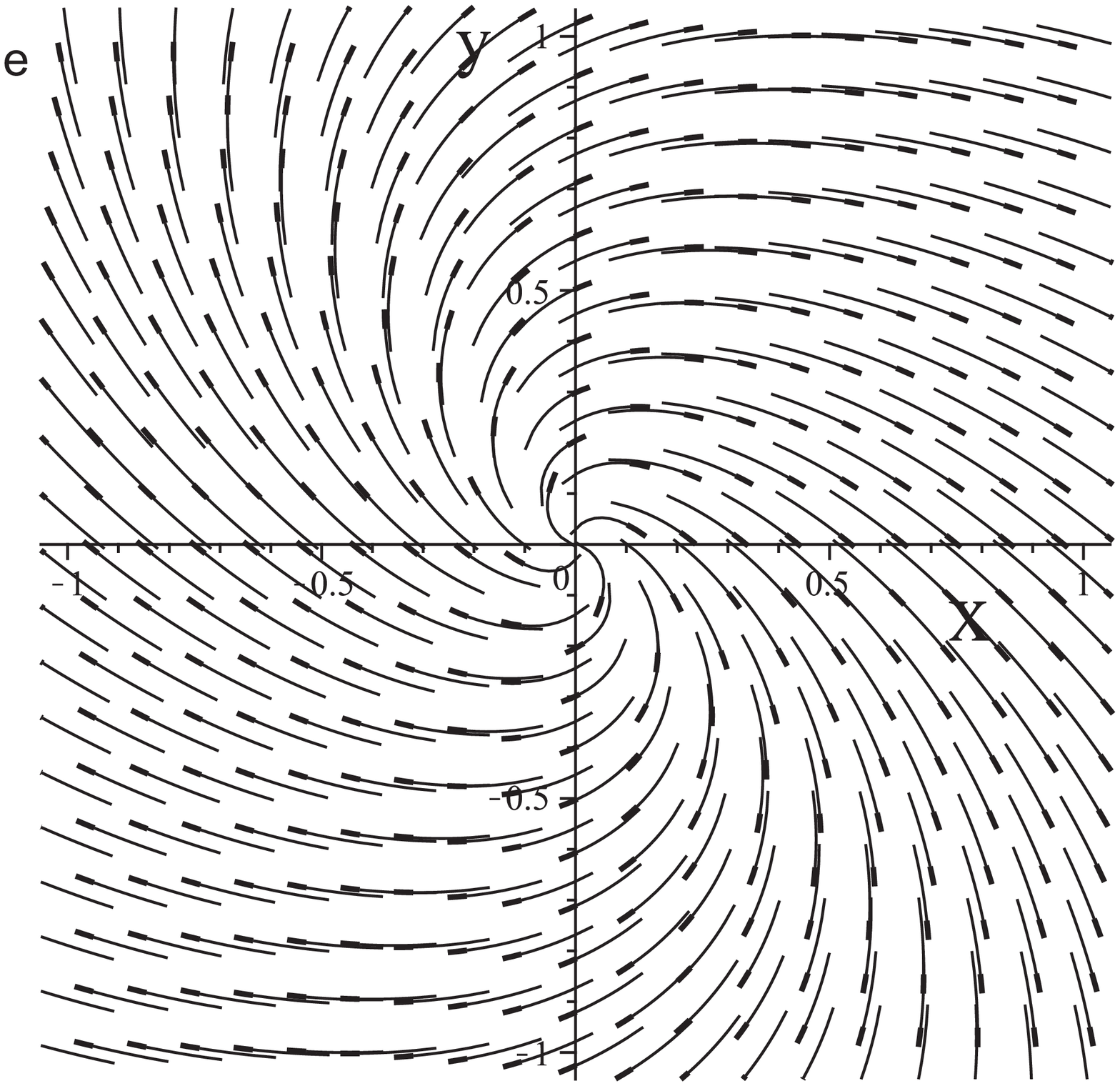}
\includegraphics[width=0.3\textwidth]{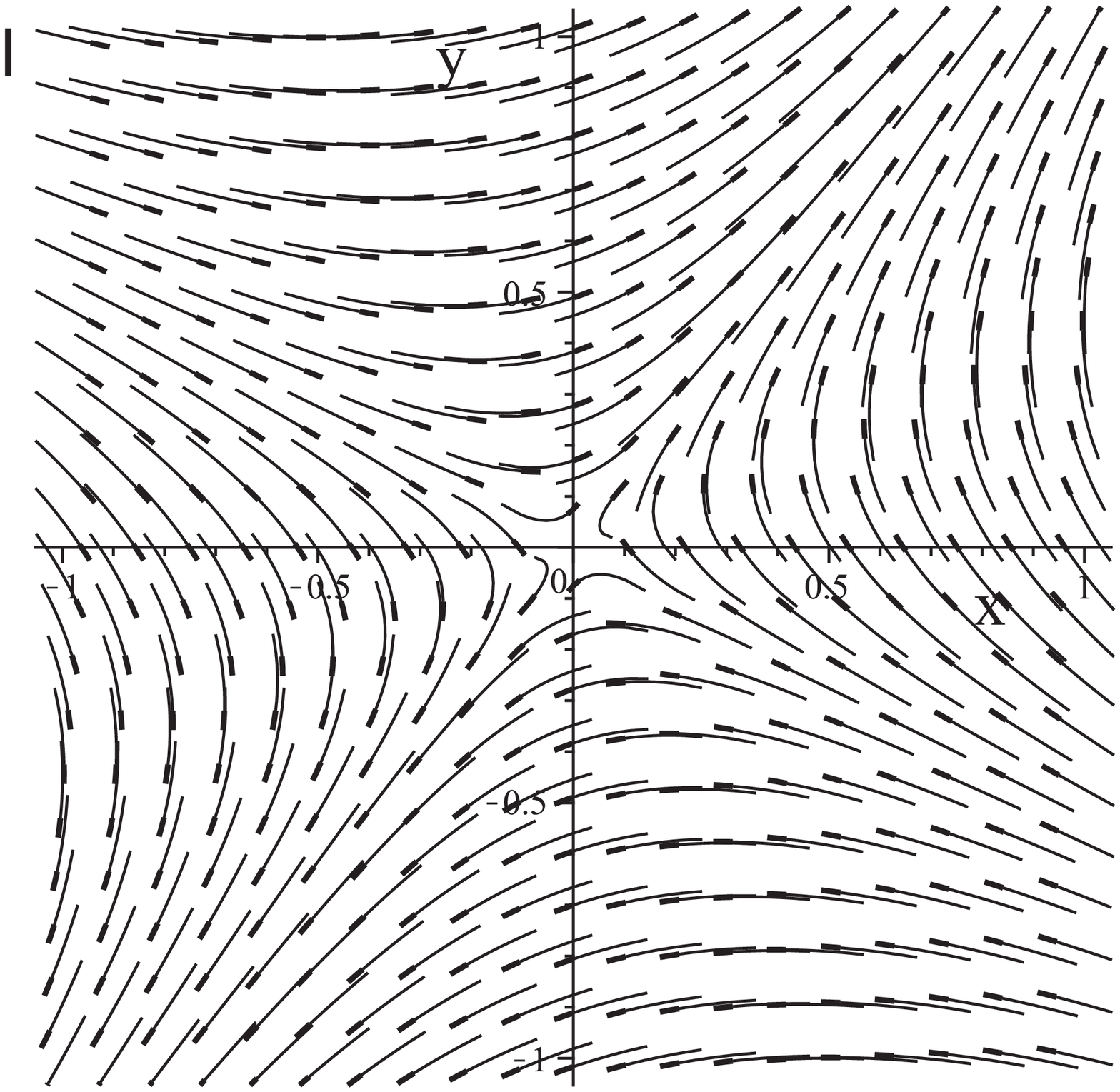}
\caption{Phase portraits for some types of critical points for $w=-1$:
a) stable node,    $\gamma =-2,\; \sigma=3,\;\beta=-1,\;\delta=3 $,
b) stable focus,   $\gamma =-3,\; \sigma=1,\;\beta=2,\;\delta=3 $,
c) center,         $\gamma =3,\; \sigma=3,\;\beta=-1,\;\delta=1 $,
d) unstable focus, $\gamma =3,\; \sigma=1,\;\beta=2,\;\delta=3 $,
e) unstable node,  $\gamma =3,\; \sigma=-3,\;\beta=2,\;\delta=3 $,
i) saddle,         $\gamma =3,\; \sigma=-3,\;\beta=1,\;\delta=-2 $,  }
\label{fig:3}
\end{figure}

There are six types of critical points:
\begin{enumerate}

 \item Stable node $\tau_{\pm}\in\Re,~\tau_{\pm}< 0, \tau_+ >\tau_- >0 $, $ \theta+\upsilon<0, ~4(\delta^2\sigma\gamma+\theta\upsilon) <(\theta+\upsilon)^2, \delta^2\sigma\gamma+\theta\upsilon>0.$

  \item  Unstable node:  $\tau_{\pm}\in\Re,~\tau_{\pm}> 0, \tau_+ >\tau_- >0 $,$ \theta+\upsilon>0, ~4(\delta^2\sigma\gamma+\theta\upsilon) <(\theta+\upsilon)^2, \delta^2\sigma\gamma+\theta\upsilon>0.$

  \item Saddle point: $\tau_{\pm}\in\Re,~ \tau_+\tau_- <0,$
 $ ~\delta^2\sigma\gamma+\theta\upsilon<0.$

  \item Stable spiral point: $\tau_{\pm}\in\mathbb{C},~ \tau\pm = \tau_1\pm i\tau_2,~\tau_1,\tau_2\in\Re~\tau_1,\tau_2>0, $ $\theta+\upsilon<0,(\theta+\upsilon)^2<4(\delta^2\sigma\gamma+\theta\upsilon).$

  \item Unstable spiral point:  $\tau_{\pm}\in\mathbb{C},~ \tau\pm = \tau_1\pm i\tau_2,~\tau_1,\tau_2\in\Re~\tau_1,\tau_2<0, $ $\theta+\upsilon>0,(\theta+\upsilon)^2<4(\delta^2\sigma\gamma+\theta\upsilon).$

 \item Elliptic fixed point $\tau_{\pm}\in\Im,~ \tau_\pm =\pm i\tau ,~\tau\in\Re, $ $\theta=\upsilon,~ \delta^2\sigma\gamma+\theta\upsilon >0.$
\end{enumerate}
     This completes a variety of critical points in the system \eqref{Omega_m_xsys}.
Some types of critical points that are typical for this system are shown in figure \ref{fig:3}.

In most cases the eigenvalues of the linearized system \eqref{Omega_m_xsys} will have eigenvalues with either positive, or negative and/or zero real parts. In these cases it is important to identify which orbits are attracted to the singular point, and which are repelled away as the independent variable (usually t) tends to infinity.

This is not a true phase-space plot, despite the superficial similarities. One important difference is that a universe passing through one point can pass  through the same point again but moving backwards along its trajectory, by first going to infinity and then turning around (recollapsing).

The local dynamics of a singular point may depend on one or more parameters. When small continuous changes in the parameter result in dramatic gripping changes in the dynamical behavior, the singular point is said to undergo a bifurcation. The values of the parameters which result in a bifurcation at the singular point can often be located by  investigating the linearized system. Singular point bifurcations will only happen if one (or more) of the eigenvalues of the linearized systems are a functions of the parameter. The bifurcations are located at the parameter values for which the real part of an eigenvalue is zero \cite{AAColey}.
The figure \ref{fig:3} actually shows such bifurcations. Different types of critical points correspond to different values ​​of parameters and hence different roots \eqref{Omega_m_xsys_roots} of the characteristic equation \eqref{Omega_m_xsys_char_eq}.

\subsection{The numerical solution for the case under interest}
Unlike the linear system of ordinary differential equations \eqref{Omega_m_xsys}, the non-linear system \eqref{Omega_xsys} allows existence of singular structures which are more intricate than set of the isolated points, fixed lines or periodic orbits.

In this subsection we consider the case which seems to us particularly interesting, mainly because it corresponds to a very interesting regime of evolution - the transient acceleration.

Starobinsky \cite{starobinsky} with co-authors, based on independent
observational data, including the brightness curves for SNe Ia,
cosmic microwave background temperature anisotropy and baryon
acoustic oscillations (BAO), were able to show, that the acceleration of Universe expansion
reached its maximum value and now decreases. In terms of the
deceleration parameter it means that the latter reached its minimum
value and started to increase. Thus the main result of the analysis
is the following: SCM is not unique though the simplest explanation
of the observational data, and the accelerated expansion of Universe
presently dominated by dark energy is just a transient phenomenon.

\begin{figure}[t]
\centering
\includegraphics[width=0.45\textwidth]{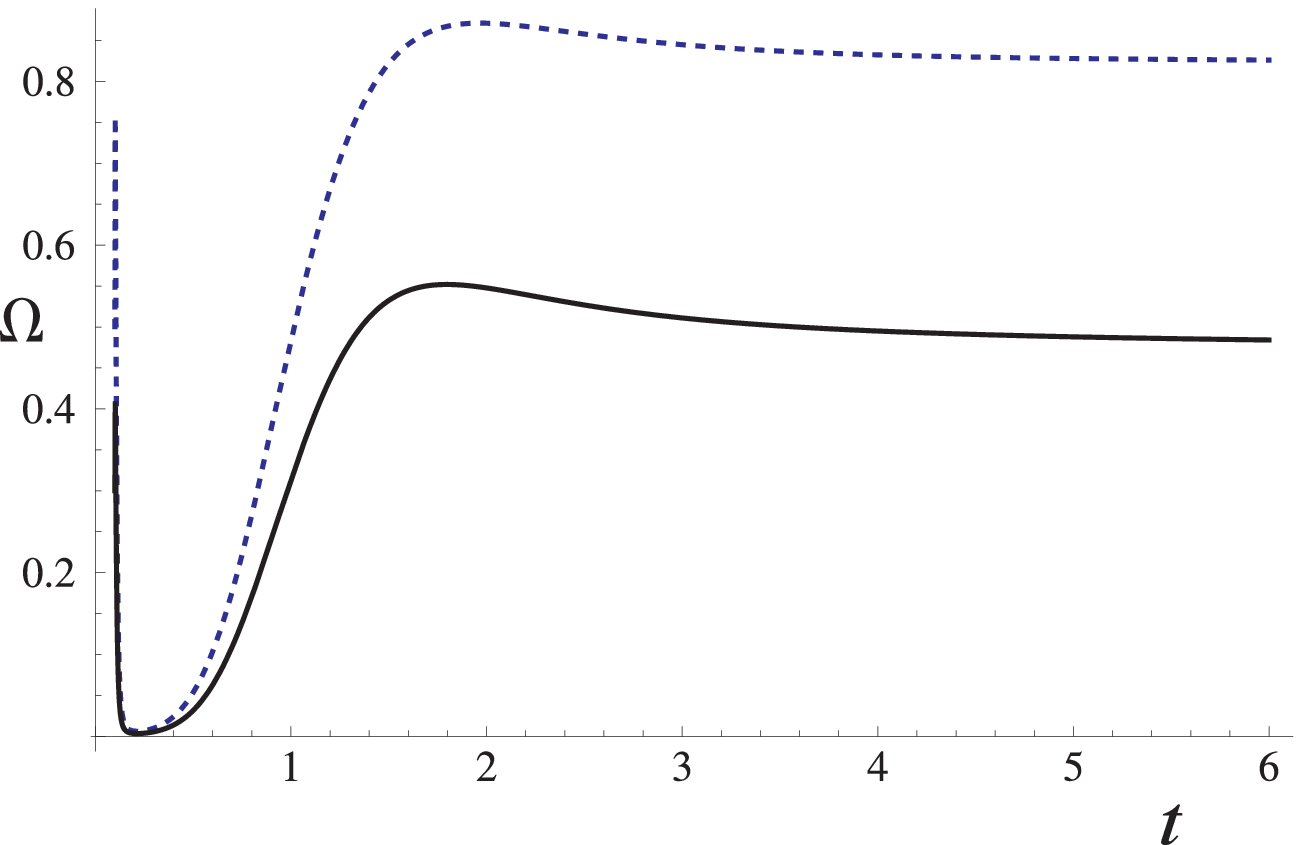}
\includegraphics[width=0.45\textwidth]{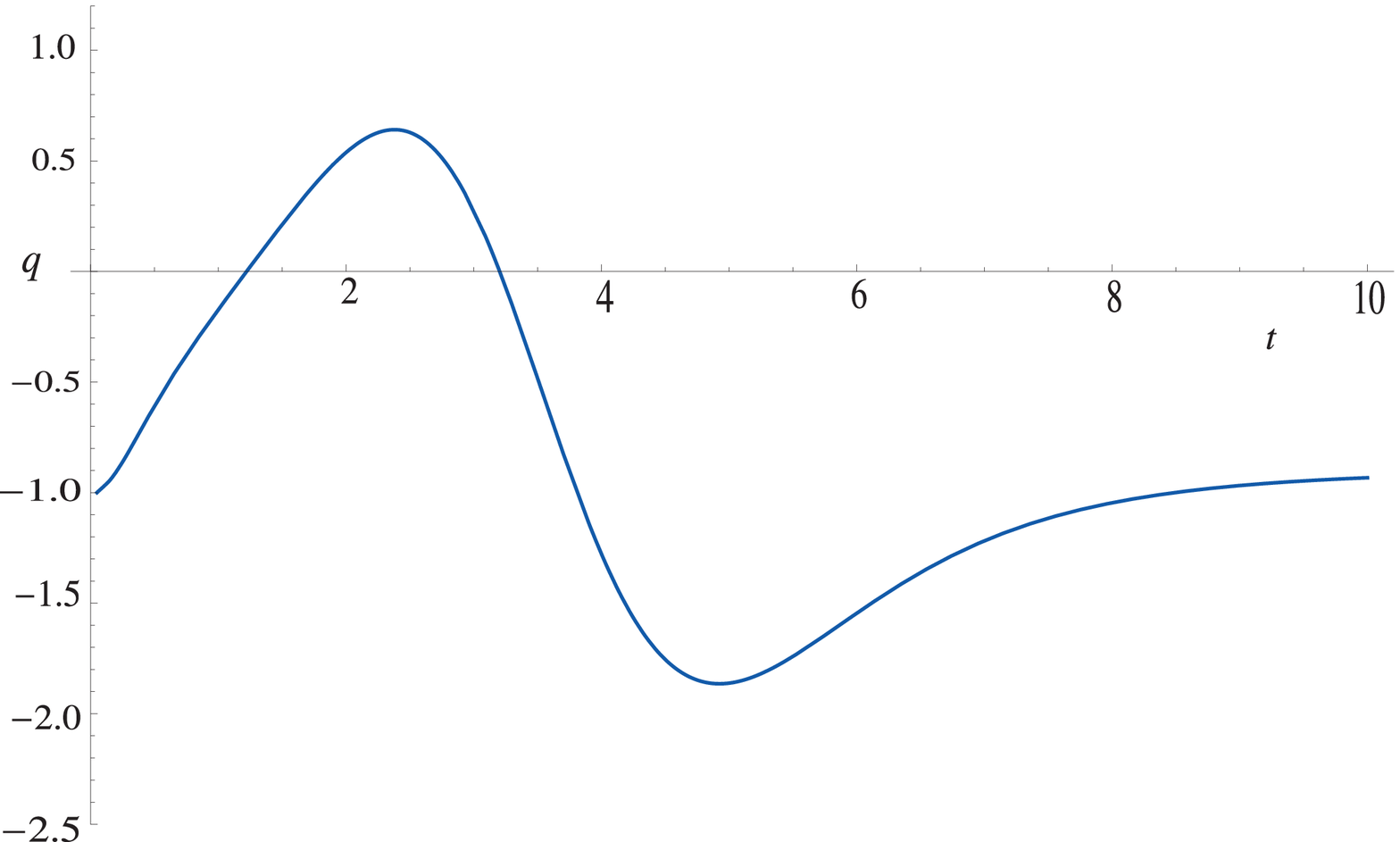}
\caption{Behavior of $\Omega_{x}$ (dot line),
$\Omega_{m}$ (solid line) (left side) and $q$ (right  side) as a function of  $t$ for $w=-1.2,\;\beta = 2,\; \delta = 1.5,\; \gamma=-2 $ and $\omega=-1$.}
\label{fig:2}
\end{figure}

Transient acceleration appears in many models of dark energy as the
interaction between dark energy and dark matter as in interaction-free models.
One of the key features of this model is that it contains only of the transient acceleration regime.

 Another feature of this model (both in interacting and interacting-free cases) is the fact that even in the early stages of the evolution of the universe, its expansion was accelerated. So, this model has a finite period of accelerated expansion (inflation) in the early stages of the Universe evolution.
 After this transition, the Universe enters a stage of slow expansion. The duration of this regime depends on values of the parameters, it is significant that it is limited. Transition period ends by phantom mode expansion, which tends asymptotically to $t \to \infty.$

The dependence of relative density of dark matter and dark energy also differs significantly in most models. Since the early times of the evolution of the universe, dominated by the fractal energy density \eqref{Rel_denFr_time}, the following holds: $\Omega_f+\Omega_m+\Omega_x=1.$ So, the contribution of the fractal  component is significant only in the early stages of the evolution of the universe.

\subsection{Interacting holographic dark energy in fractal universe}
In this section we extend our study to the case where both dark components - the pressureless
dark matter and the the Holographic model of Dark Energy (HDE) - are not conserved separately but interact with each other in fractal universe.
Based on cosmological state of holographic principle, proposed by Fischler and Susskind \cite{fischler}, HDE has been proposed and studied widely in the  literature \cite{miao,HDE}.  In HDE, in order to determine the proper and well-behaved system's IR cut-off, there are some difficulties that must be studied carefully to get results that agree
with observations that claim that our universe has accelerated expansion.

In the present section, using the holographic models of dark energy, we obtain equation of state for interacting holographic dark energy density in a fractal universe bounded by  $L$ as the system's IR cut-off.

Now we obtain the  equation for the case of interacting holographic dark energy.
In the holographic dark energy scenario, the energy density of dark energy is related to the IR cut-off $L$ by

\begin{equation}
\rho_{_L} = 3c^2M_p^2L^{-2},
\label{holo}
\end{equation}
where $c$ is a positive constant.
Differentiating the above equation with respect to time, we obtain
\begin{equation}
\dot\rho_{_L} = -2\rho_{_L}\frac{\dot L}{L}.
\label{drd1}
\end{equation}
From the definition of the relative density  \eqref{Rel_den}, one can show that
\begin{equation}
\dot{\Omega}_{_L} =- 2\Omega_{_L}\left(\frac{\dot L}{L} +\frac{\dot H}{H}\right).
\label{oma}
\end{equation}
To determine the state parameter we use the continuity equation \eqref{CeQFrm} and definition  of holographic dark energy  density \begin{equation}\label{statepar}
  w_{_L}=-1+\frac{1}{3H-\beta t^{-1}}\left(2\frac{\dot L}{L}-\frac{QL^2}{3c^2M_p^2}\right).
\end{equation}
 Here, as before,  we restrict ourselves to linear type of interaction in form \eqref{Q lin}. In this case one can obtain
\begin{equation}\label{stateparQ}
  w_{_L}=-1+\frac{1}{3H-\beta t^{-1}}\left[2\frac{\dot L}{L}-H\left(\delta + \frac{\gamma}{c^2}\frac{\Omega_m}{\Omega_{_L}}\right)\right].
\end{equation}
We note that the formula \eqref{stateparQ}  represent so-called equivalent state parameter for an interacting holographic dark energy.
The deceleration parameter, which is  defined by $q=-1-\frac{\dot{H}}{H^2}$ takes the following form
\begin{equation}\label{statepar}
  q=\frac{1}{2}((1+3w_{_L})\Omega_{_L} +\Omega_m)+\frac{\beta}{2H^2t^2}\left(\beta +1-t+\frac{2\omega\beta }{3t^{2\beta }}\right).
\end{equation}
The last term in this equation is  the distinctive feature of the fractal cosmology. Interaction parameters are not included in the expression for the deceleration parameter, and only affect the dynamic behavior of the relative densities of dark matter and dark energy.

Next, we consider various models of holographic dark energy, without regardless  the question of their viability.

As a first simple example, we consider the holographic dark energy model with Hubble horizon as an IR cut-off. In this case, the formula \eqref{holo} takes the form
\begin{equation}
\rho_{_H} = 3c^2M_p^2H^{2}.
\label{holoH}
\end{equation}
One can verify that in this case the equation of state parameter for an interacting holographic dark energy is given by
\begin{equation}\label{stateparQH}
  w_{_H}=-1-\frac{1}{3H-\beta t^{-1}}\left[2\frac{\dot H}{H}+H\left(\delta + \frac{\gamma}{c^2}\Omega_m\right)\right].
\end{equation}
It is easi to see that the accelerated expansion of the universe is possible, furthermore the phantom-accelerated regime of the universe expansion is also possible.

 Thus, in the frame of fractal cosmology within the holographic dark energy model with Hubble horizon as an IR cut-off -- the simplest and most natural model of holographic dark energy, the accelerated expansion of the universe can be explained and the coincidence problem can be solved.

As a second example we consider the holographic Ricci dark energy
(hereafter, abbreviated as RDE) model. In the RDE model, the IR
length scale $L$ is given by the average radius of the Ricci scalar
curvature ${|\cal R|}^{-1/2}$, so in this case the density of the
holographic dark energy is $\rho_{_{RDE}}\propto |{\cal R}|$. In a
spatially flat universe, the Ricci scalar of the spacetime is given
by ${\cal R}=-6(\dot{H}+2H^2)$. Therefore, the density of dark
energy can be expressed as~\cite{Gao:2007ep}
\begin{equation}
\label{rRDE} \rho_{_{RDE}}=3c^2M_p^2(\dot{H}+2H^{2}).
\end{equation}
where $c$ is a dimensionless parameter whose definition does  not coincide with \eqref{holoH}.
The relative density takes the form
\begin{equation}
\label{OmRDE} \Omega_{_{RDE}}=c^2\left(2+\frac{\dot{H}}{H^{2}}\right).
\end{equation}

To determine the state parameter, we need to find the ratio $\dot L/L.$  Taking into account the definition of IR cut-off $L = {|\cal R|}^{-1/2}$ one can  obtain
\begin{equation}
\label{OmRDE} \frac{\dot L}{L}=\frac{3c^2}{\Omega_{_{RDE}}}\left(\frac{\ddot{H}}{H^2}+4\frac{\dot H}{H}\right).
\end{equation}

The state parameter of the model under the consideration takes the following form
\begin{equation}\label{stateparOmRDE}
  w_{_{RDE}}=-1+\frac{1}{3H-\beta t^{-1}}\left[\frac{6c^2}{\Omega_{_{RDE}}}\left(\frac{\ddot{H}}{H^2}+4\frac{\dot H}{H}\right)-H\left(\delta + \frac{\gamma}{c^2}\frac{\Omega_m}{\Omega_{_L}}\right)\right].
\end{equation}

From the expression for the state parameter of this type of holographic dark energy, one can see that in the late-time stages of the universe evolution  it comes to the stage of accelerated expansion. Note that this statement is true only for the expanding universe.  Nevertheless, in the limit $t \to \infty$,  this model should not differ from conventional cosmological models, in which there is no fractal terms.
Because they contribute to the accelerated expansion of the universe that can be expected to decrease the rate of accelerated expansion. This phenomenon in the observational cosmology was called the transient acceleration \cite{starobinsky}.

\subsection{Interacting scalar field in fractal universe}

A different of theories of fundamental physics predict forecast the existence of scalar fields, motivating the study of the dynamical properties of scalar fields in cosmology. Really, scalar field cosmological models are of great importance in the study of the early universe, particularly in the investigation of inflation. Recently there has also been great interest in the late-time cosmic evolution of scalar field dark energy models.

In this section we will consider a scalar field with an exponential potential. Models with a self-interaction potential with an exponential dependence on the scalar field of the form

$$V(\phi)=V_0\exp(-\frac{\lambda}{M_p}\phi)$$ evolving in a spatially-flat FRW fractal universe and interacting with the matter. This potential where $V_0$ and $\lambda$ are positive constants.
The scalar potential in usual (non-fractal) cosmology without interaction, acts as a cosmological constant.  Exponential potentials emerge very ordinarily in all models of unification with gravity as Kaluza-Klein theories, supergravity theories or string theories. In higher dimensional theories the
scalar could be associated with the volume of “internal space”. The exponential
form of the potential reflects here the fact that time derivatives in gravity typically
involve the logarithm of length scales.

 The equations (\ref{cont_eq}) for matter and scalar field transform to
\begin{eqnarray}
  &&\dot{\rho}_m+\left(3H-\beta t^{-1}\right)\rho_m=Q,\label{Sc_m}\\
 &&\ddot{\phi}+(3H-\beta t^{-1})\dot{\phi}+\frac{dV}{d\phi}=-\frac{Q}{\dot{\phi}}.\label{KG}
\end{eqnarray}
We consider such coupling that
$$Q=\sqrt{\frac{2}{3}}\frac{\gamma\rho_m\dot{\phi}}{M_p}.$$
Such a coupling arises for instance in string theory, or after a conformal transformation of Brans-Dicke theory \cite{Wetterich_1995}, \cite{Amendola_99}.

In this case it is convenient to introduce the following variables
\begin{equation}\label{xyzuv}
x^2 = \frac{\dot{\phi}^2}{6M_p^2H^2},~~ y^2 = \frac{V}{3M_p^2H^2},~~ z^2 = \frac{\rho_m}{3M_p^2H^2},~~ u^2 = \frac{\omega\beta^2}{6H^2t^{2(\beta+1)}},~~ s^2 = \frac{\beta}{Ht}.
\end{equation}
Note that in this new variables the relative densities of the components take the form
\begin{equation}\label{omega_xyzuv}
\Omega_x = x^2+y^2,~~ \Omega_m= z^2,~~ \Omega_f=u^2+ s^2,
\end{equation}
where $\Omega_f$ is fractal relative density as mentioned above. Then Friedman equation becomes
\begin{equation}\label{scalar_frid}
    1=\Omega_x+\Omega_m+\Omega_f=x^2+y^2+z^2+u^2+s^2
\end{equation}

The evolution equations can then be written as a plane autonomous system:
\begin{eqnarray}
  &&x'=-x\left[\frac{H'}{H}+3-s^2\right]+\sqrt{\frac 32}\lambda y^2-\gamma z^2,\label{scalar_x}\\
 &&y'=-y\left[\frac{H'}{H}+\sqrt{\frac 32}\lambda x\right],\label{scalar_y}\\
 &&z'=-z\left[\frac{H'}{H}+\frac 12(3-s^2)-\gamma x\right],\label{scalar_z}\\
 &&u'=-u\left[\frac{H'}{H}+s^2\left(1+\frac1\beta\right)\right],\label{scalar_u}\\
 &&s'=-\frac12 s\left[\frac{H'}{H}+\frac{s^2}{\beta}\right],\label{scalar_v}\\
 &&\frac{H'}{H}=-1+\frac12\left[s^2-z^2-s^4\left(1-\frac1\beta\right)\right]-2u^2-2x^2+y^2
\end{eqnarray}
where the prime denotes  $d/dN$ and $N=\ln a$.

First of all, it should be noted that the definition of the variable $u^2$ includes the parameter $\omega$, which may be negative, so  unlike others the variable $u^2$ can also take a negative value .

There are no critical points in this multi-dimensional system in proper sense, but there are critical lines, periodic orbits and surfaces. It differs significantly from similar systems in cosmology.
In this case, it is extremely easy to obtain the relation
$$x=\frac{3}{2\gamma}-\frac{s^2}{\gamma}\left(\frac{1}{2}+\frac{1}{\beta}\right),$$
which holds for the critical case and $u \to \infty .$ For $s = 0$ (no fractal components), we obtain the standard result.

\section{Conclusions}
In the present paper we consider a new model of interacting dark energy and dark matter with additional time dependent term within the framework of fractal cosmology. This model demonstrates new types of evolution, which are not common to cosmological models with this type of interaction. Within this model cosmological parameters depend on time in a strongly nonlinear manner. In particular, in some solutions the deceleration parameter is a non-monotonic function of time, revolving between the stages of accelerated and decelerated expansion, thus demonstrating transient acceleration.  The numerical solutions for some interesting cases are also shown.

We also examined both of interacting holographic dark energy and dark energy models in the form of a scalar field in the context of fractal cosmological model. At a qualitative level, it is shown that in these models there are solutions that allow to solve some cosmological problems.

In all the investigated models we found that stable late-time solutions do exist, corresponding both to accelerated and slowed down regimes of universe evolution (this can be true on different time scales).  Indeed, for all the studied models, we did not find any non-accelerating stable solution. However, in almost all the cases the late-time solutions correspond to the dark energy domination and thus are unable to solve the coincidence problem, that, in principle, characterizes most models of interacting components.

\acknowledgments

We are grateful to   Prof. Yu.L. Bolotin for kind help and discussions. We also thank V.A. Cherkaskiy  for careful reading and editing and leave time for more than one draft before final submission.
Work is supported in part by the Joint DFFD-RFBR Grant \# F40.2/040.

\end{document}